\documentclass[conference]{IEEEtran}

\usepackage{color}

\usepackage[justification=centering]{caption}
\usepackage{subfigure}
\usepackage{graphicx,cite,epsfig,amssymb,amsmath,multirow,lettrine,flushend,extarrows}
\usepackage{algorithm}
\usepackage{algorithmic}

\begin{document}

\title{Optimal Power Allocation for Minimizing Outage Probability of UAV Relay Communications}

\author{\IEEEauthorblockN{Hui~Tu, Jia Zhu*, and Yulong Zou*}

\IEEEauthorblockA{School of Telecomm. and Inform. Eng., Nanjing Univ. Posts \& Telecomm., Nanjing, P. R. China}
\IEEEauthorblockA{*Corresponding authors (\{jiazhu, yulong.zou\}@njupt.edu.cn)}
}

\maketitle

\begin{abstract}
Unmanned aerial vehicle (UAV) networks have grown rapidly in recent years and become attractive for various emergence communications scenarios. In this paper, we consider a UAV acting as a relay node to assist wireless transmissions from a base station (BS) to a ground user (GUs). A closed-form expression of outage probability for the BS-GUs transmission via UAV relaying is derived over Rician fading channels. We then formulate an optimization problem to minimize the outage probability of UAV relay communications with a constraint on the total transit power of the BS and GUs. It is proved that our formulated optimization problem is convex and an optimal power allocation solution is found for the outage probability minimization. Simulation results demonstrate that with an increasing power allocation factor, the outage probability initially decreases and then starts to increase, showing the existence of an optimal power allocation solution. Additionally, it is shown that the proposed optimal power allocation scheme significantly outperforms the conventional equal power allocation in terms of the outage probability.
\end{abstract}

\begin{IEEEkeywords}
UAV relaying, Rician fading, Power allocation, Outage probability.
\end{IEEEkeywords}

\section{Introduction}

\IEEEPARstart{I}{n} various emergency communication scenarios (e.g., disaster areas, ocean areas, etc.), traditional cellular wireless networks are limited to a certain extent and can not provide the network coverage for some complex terrains. To this end, UAVs are emerging as an effective means of establishing the fast and flexible network deployment due to its strong mobility [1]. Moreover, incorporating a UAV as a relay node in existing wireless network architecture is capable of extending network coverage as well as improving the transmission throughput [2]. In UAV relay-aided networks, it is of interest to examine the optimal deployment of UAVs according to different quality-of-service (QoS) requirements. In [3], the authors proposed a mathematical model to optimize the UAV altitude that maximizes the network coverage. The impact of the UAV altitude on the network coverage was further studied in [4] by jointly considering the path loss and fading.

In addition to optimizing the UAV altitude, the deployment of its horizontal position was investigated in [5] for the sake of reducing the number of UAV base stations. In [6], the authors studied a UAV enabled multiuser communication system equipped with a directional antenna of adjustable beamwidth and proposed a joint UAV's flying altitude and antenna beamwidth optimization for throughput enhancement. In [7], the power control optimization was investigated for device-to-device (D2D) communications underlaying UAV-assisted wireless systems, where the difference of two convex functions (D.C.) programming is utilized to solve the formulated optimization problem. Moreover, the authors of [8] examined the problem of user-demand-based UAV assignment for hotspot areas with high traffic demands and proposed a so-called neural-based cost function approach for better load balancing and traffic offload.

Recently, cooperative relaying technology has received extensive attention due to its advantage of extending network coverage and improving system  capacity [10]-[12]. In [13], several cooperative relaying protocols, namely the fixed relaying, selection relaying, and increment relaying, were analyzed over Rayleigh fading channels in terms of the outage probability. The authors of [14] compared three cooperative relaying schemes, i.e., the amplify-and-forward (AF), decode-and-forward (DF) and coded cooperation (CC), showing that the outage performance of DF and CC methods are generally better than that of the AF approach. It is pointed out that the AF has a simpler implementation complexity than the DF and CC, since it just forwards an amplified version of the received signal without any sort of decoding.

In UAV communications, a line-of-sight (LoS) propagation path is available for an air-to-ground (A2G) channel, for which the Rician model [15] is preferred to characterize the small-scale fading. As a consequence, this paper considers a UAV aided wireless system, where a ground user (GU) is out of the coverage of its base station (BS) and a UAV is employed as a relay node to assist the transmission from BS to GU in the DF manner. More specifically, BS transmits its signal to a UAV that decodes its received signal and then forwards its decoded outcome to GU. The main contributions of this paper are summarized as follows. First, a closed-form expression of outage probability is derived for the UAV assisted BS-GU transmission over Rician fading channels. Second, an outage probability minimization problem is formulated with a constraint on the total transit power of the BS and GU. Finally, an optimal power allocation solution is found for minimizing the outage probability of the UAV assisted BS-GU transmission.

The rest of this paper is organized as follows. Section II describes the system model of UAV-assisted wireless systems. In Section III, we derive a closed-form outage probability expression for UAV-assisted wireless transmissions over Rician fading channels and propose an optimal power allocation scheme to minimize the derived outage probability. Next, Section IV provides numerical outage probability results to show the advantage of proposed optimal power allocation scheme. Finally, some concluding remarks are drawn in Section V.

\section{System Model}
We consider a dual-hop decode-and-forward (DF) relaying system in a downlink scenario, as shown in Fig. 1, where ground users (GUs) are out of the coverage of BS and a UAV node acts as a relay to assist the transmission from BS to GUs. To be specific, the BS transmits its signal to the UAV in the first phase. Then, in the second phase, the UAV attempts to decode its received signal from BS and forwards the decoded result to GUs. In our system, all the network nodes including the BS, UAV relay and GUs are considered to be equipped with a single antenna. For notational convenience, let $L$ denote the horizontal distance from the BS to a GUs. The horizontal distance between the BS and the UAV relay is denoted by $r_s$ and the horizontal distance between the UAV relay and a GUs destination is denoted by $r_d$, wherein ${r_s} + {r_d} = L$. Moreover, the vertical altitude of the UAV relay is represented by $h_u$.

\begin{figure}
  \centering
  {\includegraphics[scale=0.3]{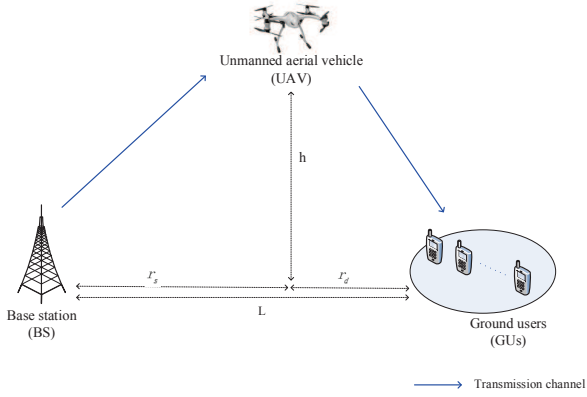}\\
  \caption{Illustration of a UAV relay network.}\label{fig1}}
\end{figure}

Considering that the LoS propagation component is often available for an A2G channel, we adopt the Rician model to characterize the fading process from BS and UAV relay and that from UAV relay to GUs. According to [3], the LoS path loss $G^{\textrm{LoS}}_{su}$ from BS to UAV and the non-line-of-sight (NLOS) path loss $G^{\textrm{NLoS}}_{su}$ from UAV and a GUs destination can be written as
\begin{equation}
G^{\textrm{LoS}}_{su} = \frac{{{c^2}d_{su}^{ - n}}}{{{{(4\pi {f_c})}^2}}}{2^{\frac{{10}}{{\eta^{\textrm{LoS}}_{su}}}}} ,
\end{equation}
and
\begin{equation}
G^{\textrm{NLoS}}_{su} = \frac{{{c^2}d_{su}^{ - n}}}{{{{(4\pi {f_c})}^2}}}{2^{\frac{{10}}{{\eta^{\textrm{NLoS}}_{su}}}}},
\end{equation}
where $c$ is the speed of light, ${f_c}$ is the carrier frequency, $d_{su}$ is the propagation distance from BS to UAV as given by $d_{su} = \sqrt {{h_u^2} + {r_s}^2}$, $n$ is the path loss exponent, and $\eta^{\textrm{LoS}}_{su}$ and $\eta^{\textrm{NLoS}}_{su}$ are excessive path losses of the LoS and NLoS propagation from BS to UAV, respectively. Similarly, the LoS path loss $G^{\textrm{LoS}}_{ud}$ from UAV to GUs destination and the NLOS path loss $G^{\textrm{NLoS}}_{ud}$ from UAV and GUs destination are expressed as
\begin{equation}
G^{\textrm{LoS}}_{ud} = \frac{{{c^2}d_{ud}^{ - n}}}{{{{(4\pi {f_c})}^2}}}{2^{\frac{{10}}{{\eta^{\textrm{LoS}}_{ud}}}}} ,
\end{equation}
and
\begin{equation}
G^{\textrm{NLoS}}_{ud} = \frac{{{c^2}d_{ud}^{ - n}}}{{{{(4\pi {f_c})}^2}}}{2^{\frac{{10}}{{\eta^{\textrm{NLoS}}_{ud}}}}},
\end{equation}
where $d_{ud}$ is the propagation distance from UAV to GUs destination as given by $d_{ud} = \sqrt {{h_u^2} + {r_d}^2}$, and $\eta^{\textrm{LoS}}_{ud}$ and $\eta^{\textrm{NLoS}}_{ud}$ are excessive path losses of the LoS and NLoS propagation from UAV to GUs destination, respectively. Following [3], the probability of LoS propagation from BS to UAV and the corresponding NLoS propagation probability are given by
\begin{equation}
P^{\textrm{LoS}}_{su} = \frac{1}{{1 + {a_{su}}\exp [ - {b_{su}}(\theta_{su} - {a_{su}})]}},
\end{equation}
and
\begin{equation}
P^{\textrm{NLoS}}_{su} = 1 - P^{\textrm{LoS}}_{su},
\end{equation}
where ${\theta _{su}} = \arctan (\frac{h_u}{{{r_s}}})$, and ${{a_{su}}}$ and ${b_{su}}$ are environment-dependant parameters for the BS-UAV channel. Moreover, the probabilities of LoS and NLoS propagation from UAV to GUs destination are written as
\begin{equation}
P^{\textrm{LoS}}_{ud} = \frac{1}{{1 + {a_{ud}}\exp [ - {b_{ud}}(\theta_{ud} - {a_{ud}})]}},
\end{equation}
and
\begin{equation}
P^{\textrm{NLoS}}_{ud} = 1 - P^{\textrm{LoS}}_{ud},
\end{equation}
where ${\theta _{ud}} = \arctan (\frac{h_u}{{{r_d}}})$, and ${{a_{ud}}}$ and ${b_{ud}}$ are environment-dependant parameters for the UAV-GUs channel. Therefore, the large-scale path loss from the BS to UAV denoted by ${G_{su}}$ and that from the UAV and GUs destination denoted by ${G_{ud}}$ can be expressed as
\begin{equation}
{G_{su}} = G^{\textrm{LoS}}_{su}P^{\textrm{LoS}}_{su} + G^{\textrm{NLoS}}_{su}P^{\textrm{NLoS}}_{su},
\end{equation}
and
\begin{equation}
{G_{ud}} = G^{\textrm{LoS}}_{ud}P^{\textrm{LoS}}_{ud} + G^{\textrm{NLoS}}_{ud}P^{\textrm{NLoS}}_{ud}.
\end{equation}

In general, an A2G channel experiences both the large-scale path loss and small-scale fading, thus the received power at UAV from BS is given by
\begin{equation}
{P_{su}} = {P_s}{G_{su}}{| {{h_{su}}} |^2},
\end{equation}
where $P_s$ is the transmit power of BS and ${h_{su}}$ is a fading coefficient of the channel from BS to UAV. Similarly, the received power at the GUs destination from the UAV relay is expressed as
\begin{equation}
{P_{ud}} = {P_u}{G_{ud}}{| {{h_{ud}}} |^2},
\end{equation}
where ${P_u}$ is the transmit power of the UAV relay and ${h_{ud}}$ is a fading coefficient of the channel from UAV to GUs destination.

The instantaneous signal-to-noise ratio between the BS and the UAV relay and UAV relay to the UG can be written as
\begin{equation}
{\gamma _{su}} = \frac{{{P_s}{G_{su}}{{\left| {{h_{su}}} \right|}^2}}}{{{N_0}}},
\end{equation}
\begin{equation}
{\gamma _{ud}} = \frac{{{P_u}{G_{ud}}{{\left| {{h_{ud}}} \right|}^2}}}{{{N_0}}},
\end{equation}
where $N_0$ denotes the noise power.

In this paper, the Rician distribution is used to model the small-scale fading. Using the Rician model [9], the non-central chi-square probability distribution function (PDF) is adopted to characterize the distribution of ${\gamma _{su}}$ and ${\gamma _{ud}}$, namely
\begin{equation}
{f_\gamma }\left( x \right) = \frac{{\left( {K + 1} \right)}}{{{e^K}\overline \gamma  }}\exp \left[ {\frac{{ - (K + 1)x}}{{\overline \gamma  }}} \right] {I_0}\left[ {2\sqrt {\frac{{K(K + 1)x}}{{\overline \gamma  }}} } \right],
\end{equation}
for $x \ge 0$, where $\overline \gamma   = E\left( \gamma  \right)$, ${I_0}( \cdot )$ is the zero-order modified Bessel function, and $K$ is the Rician factor which is a function of the elevation angle. According to [4], the Rician factor is given by $K = a{e^{b\theta }}$, where $a$ and $b$ are environment dependant parameters as given by $a = {K_0}$ and $b = \frac{2}{\pi }\ln (\frac{{{K_{{\pi /2}}}}}{{{K_0}}})$, in which ${K_0}$ is the minimum value of $K$ for an elevation angle of $\theta  = 0$ and ${K_{\pi/2}}$ is the maximum value of $K$ for an elevation angle of $\theta  = \pi / 2$. Typically, as the elevation angle $\theta $ increases, the Rician factor and the LOS probability increase accordingly, resulting in a lower shadow and diffraction effect.

According to [10]-[12], an instantaneous channel capacity from the BS to UAV relay is given by
\begin{equation}
{C _{su}} =\frac{1}{2}\log_2( 1 + \frac{{{P_s}{G_{su}}{{| {{h_{su}}} |}^2}}}{{{N_0}}}),
\end{equation}
where $G_{su}$ is given by (9). Moreover, an instantaneous channel capacity from the UAV relay to GUs destination is expressed as
\begin{equation}
{C _{ud}} =\frac{1}{2}\log_2( 1 + \frac{{{P_u}{G_{ud}}{{| {{h_{ud}}} |}^2}}}{{{N_0}}}),
\end{equation}
where $G_{ud}$ is given by (10). Noting that the DF protocol is adopted at the UAV relay, we obtain that an overall channel capacity from the BS via UAV relay to GUs is the minimum of $C_{su}$ and $C_{ud}$, namely
\begin{equation}
{C _{sd}}  = \min \left( {C _{su}}, {C _{ud}} \right),
\end{equation}
where ${C _{su}}$ and ${C _{ud}}$ are given by (14) and (15), respectively.

\section{Optimal Power Allocation for Outage Probability Minimization}
In this section, we first derive a closed-form expression of the outage probability for the UAV relay transmission and then propose an optimal power allocation scheme to minimize the derived outage probability. Following [10]-[12], an outage probability of the UAV relay transmission is obtained as
\begin{equation}
{P_{\textrm{out}}} = \Pr\left( {C_{sd}  < R} \right) ,
\end{equation}
where $R$ is a data rate of the BS-GUs transmission. Substituting $C_{sd}$ from (16) into (17) gives
\begin{equation}
{P_{\textrm{out}}} = \Pr\left[ {\min \left( {C _{su}},{C _{ud}} \right) < R} \right] \\.
\end{equation}
Substituting $C_{su}$ and $C_{ud}$ from (16) and (17) into the preceding equation yields
\begin{equation}
\begin{split}
{P_{\textrm{out}}}& = \Pr\left[ {\min \left( {C _{su}},{C _{ud}} \right) < R} \right]   \\
& = 1 - {Q_1}\left( {\sqrt {2{K_{su}}} ,\sqrt {2({K_{su}} + 1){\gamma _{a}}} } \right)  \\
&\quad\quad \times {Q_1}\left( {\sqrt {2{K_{ud}}} ,\sqrt {2({K_{ud}} + 1){\gamma _{b}}} } \right),
\end{split}
\end{equation}
where ${\gamma _{a}} = \frac{\left( {{2^{2R}} - 1} \right){N_0}}{{{P_s}{G_{su}}}}$, ${\gamma _{b}} = \frac{\left( {{2^{2R}} - 1} \right){N_0}}{{{P_u}{G_{ud}}}}$, ${Q_1}( \cdot, \cdot )$ is the first order Marcum Q-function, and ${K_{su}}$ and ${K_{ud}}$ represent a Rician factor of the channel from the BS to the UAV and that from the UAV to the GU, respectively.

Our goal is to minimize the outage probability through an optimization of the power allocation between the BS and UAV, which is described as
\begin{equation}
\begin{split}
&\mathop {\min }\limits_{{P_s},{r_s},{h}} {P_{\textrm{out}}}, \\
&{\textrm{s.t.}}\quad {P_s} + {P_u} \le {P_t}, \\
& \quad \quad {P_s} \ge 0,{P_d} \ge 0,
\end{split}
\end{equation}
where $P_{\textrm{out}}$ is given by (21). Since the outage probability of proposed optimal power allocation scheme monotonically decreases with an increasing ${P_s}$ and ${P_u}$. Thus, an increase in the total transmit power decreases the outage probability. This means that a minimized outage probability is obtained when ${P_s} + {P_d} = {P_t}$ is satisfied.

\textbf{\emph{Theorem 1:}}
\emph{For the distance constrained, the approximate expression of the optimal power is written as
\begin{align} \label{f_11}
&{\left( {\frac{{{P_u}}}{{{P_s}}}} \right)^{\frac{7}{4}}}{\left( {\frac{{{\gamma _1}}}{{{\gamma _2}}}} \right)^{\frac{3}{4}}}\sqrt {\frac{{{K_{ud}}}}{{{K_{su}}}}} \exp [2\sqrt {\frac{{{\gamma _1}}}{{{P_s}}}}  - 2\sqrt {\frac{{{\gamma _2}}}{{{P_u}}}} \\
& \qquad\qquad\qquad\qquad\qquad\qquad + {K_{ud}} - {K_{su}}] = 1,\\ \notag
\end{align}
where \\
${\gamma _1} = \frac{{{K_{su}}({K_{su}} + 1){({{2^{2R}} - 1})}{N_0}}}{{{G_{su}}}}, {\rm{ }}{\gamma _2} = \frac{{{K_{ud}}({K_{ud}} + 1){({{2^{2R}} - 1})}{N_0}}}{{{G_{ud}}}}$.}
\begin{IEEEproof}
See Appendix A.
\end{IEEEproof}

\section{Numerical Results and Discussions}
In this section, we present numerical outage probability results of the proposed optimal power allocation and conventional equal power allocation schemes. According to the existing literature [], the adopted system parameters are shown in Table I. For notational convenience, a power allocation factor $\alpha$ ($0 \le \alpha \le 1$) is used to represent the ratio of the BS' transmit power $P_s$ to the total transmit power $P_t$, leading to $P_s = \alpha P_t$ and $P_u = (1 - \alpha)P_t$. Moreover, the UAV relay is deployed at the center of the BS and GU destination, i.e., $r_s = r_d = L/2$. In addition, a total transmit power of $P_t = 0.25w$, a data rate of $R=1$bit/s/Hz, the UAV's altitude of $h=1000$ meters ($m$), and BS-GU distance of $L=2000m$ are considered in our numerical evaluation, unless otherwise stated.

\begin{table}[!h]
\caption{System Parameters}
\centering
\begin{tabular}{|c|c|c|}
\hline
Simulation Parameters & Value\\
\hline
Carrier frequency (${f_c}$) & 2000MHz \\
\hline
Excessive pathloss in BS $\left( {\eta^{\textrm{LoS}}_{su},\eta^{\textrm{NLoS}}_{su}} \right)$ & $(1,20)$dB \\
\hline
Excessive pathloss in GU $\left( {\eta^{\textrm{LoS}}_{ud},\eta^{\textrm{NLoS}}_{ud}} \right)$ & $({\rm{1}}{\rm{.6}},2{\rm{3}})$dB \\
\hline
BS-UAV channel parameters $\left( {{a_{su}},{b_{su}}} \right)$ & $(0.28,9.6)$dB \\
\hline
UAV-GU channel parameters $\left( {{a_{ud}},{b_{ud}}} \right)$ & $(0.136,11.95)$dB \\
\hline
Total transmit power (${P_t}$) & 0.25$w$ \\
\hline
Rician factors $\left( {{K_0},{K_{\pi/2}}} \right)$ & $(5,15)$dB \\
\hline
Data rate ($R$) & $1$ bit/s/Hz \\
\hline
Path loss exponent ($n$) & 3 \\
\hline
Noise power (${N_0}$) & -110dBm \\
\hline
\end{tabular}
\end{table}
\begin{figure}[htbp]
\begin{centering}
\includegraphics[scale=0.6]{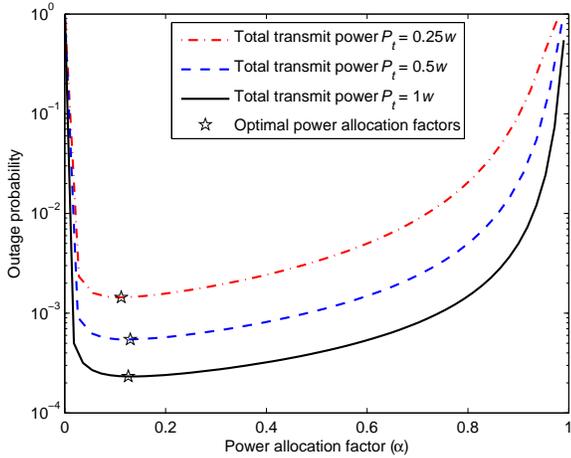}
\par\end{centering}
\caption{Outage probability versus power allocation factor $\alpha$ of UAV relay transmissions for different total transmit powers $P_t$.}\label{fig2}
\end{figure}
Fig. 2 shows the outage probability versus power allocation factor $\alpha$ of the UAV relay transmission for different total transmit powers $P_t$, where the outage probability curves are plotted by using (19) and the optimal power allocation factors are obtained by (21). As can be seen from Fig. 2, as the power allocation factor increases, the outage probability of UAV relay transmissions initially decreases and then starts to increase, showing the existence of an optimal power allocation factor in terms of minimizing the outage probability. Moreover, Fig. 2 shows that the optimal power allocation factors match well with the minimum values of their corresponding outage probability curves, verifying the correction of our optimal power allocation solution as given by (21). One can also see from Fig. 2 that as the total transmit power increases from $P_t = 0.25w$ to $1w$, the outage probability of UAV relay transmissions improves significantly.

\begin{figure}[htbp]
\includegraphics[scale=0.6]{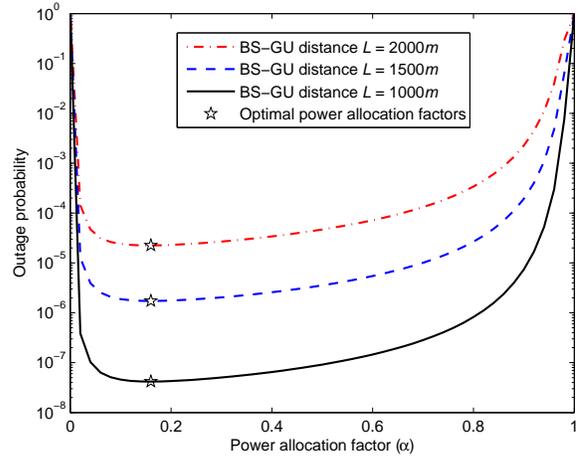}
\caption{Outage probability versus power allocation factor $\alpha$ of UAV relay transmissions for different BS-GU distances $L$.}\label{fig3}
\end{figure}

Fig. 3 depicts the outage probability versus power allocation factor $\alpha$ of UAV relay transmissions for different BS-GU distances $L$. It is shown from Fig. 3 that for any cases of $L=1000m$, $1500m$ and $2000m$, the outage probability first decreases to the minimum and then begins to increase, further validating the existence of an optimal power allocation factor for the outage probability minimization. Fig. 3 also shows that as the BS-GU distance increases from $L=1000m$ to $2000m$, the outage probability of UAV relay transmissions increases. This is due to the fact that with an increasing BS-GU distance, a higher path loss is encountered and thus the received signal power at GU weakens, leading to an increased outage probability. Additionally, one can observe from Fig. 3 that the optimal power allocation factors match well with the minimum values of outage probability curves.

\begin{figure}[htbp]
\centering
\includegraphics[scale=0.6]{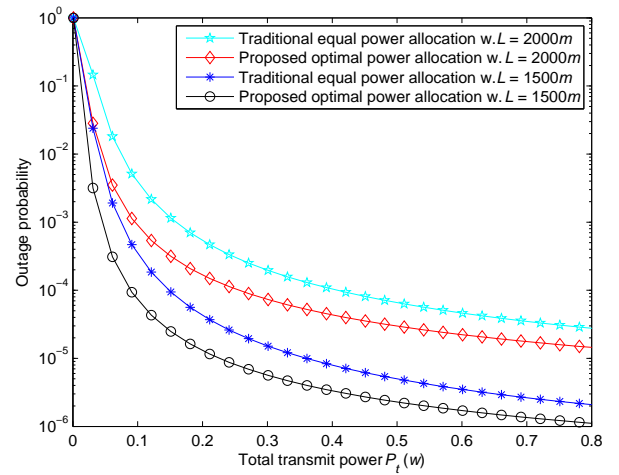}
\caption{Outage probability versus the total transmit power $P_t$ of the proposed optimal power allocation and conventional equal power allocation schemes for different BS-GU distances $L$.}\label{fig4}
\end{figure}

Fig. 4 illustrates the outage probability versus the total transmit power $P_t$ of the proposed optimal power allocation and conventional equal power allocation schemes for different BS-GU distances $L$. As shown in Fig. 4, for both cases of $L=1500m$ and $2000m$, the outage probabilities of proposed optimal power allocation and conventional equal power allocation schemes decrease monotonically with an increase of the total transmit power. Moreover, as the BS-GU distance decreases from $L=2000m$ to $1500m$, the outage probability of UAV relay transmissions decreases accordingly. Fig. 4 also shows that for both cases of $L=1500m$ and $2000m$, the proposed optimal power allocation scheme significantly outperforms the conventional equal power allocation method.

\begin{figure}[htbp]
\centering
\includegraphics[scale=0.6]{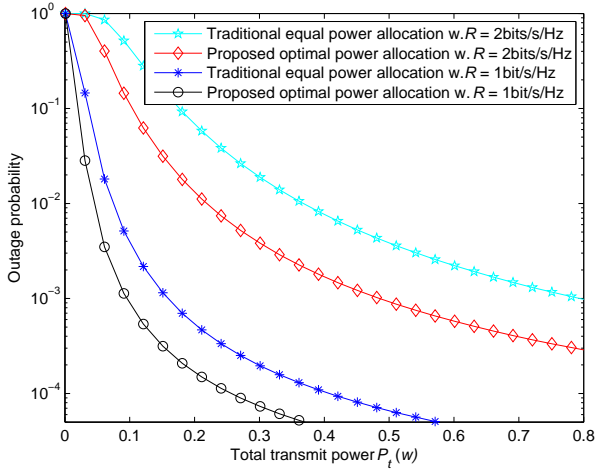}
\caption{Outage probability versus the total transmit power $P_t$ of the proposed optimal power allocation and conventional equal power allocation schemes for different data rates $R$.}\label{fig4}
\end{figure}

Fig. 5 shows the outage probability versus the total transmit power $P_t$ of the proposed optimal power allocation and conventional equal power allocation schemes for different data rates $R$. One can see from Fig. 5 that as the data rate increases from $R = 1$bit/s/Hz to $2$bits/s/Hz, the outage probabilities of proposed optimal power allocation and conventional equal power allocation schemes both increase. In addition, the outage probability of proposed optimal power allocation scheme is shown to be much smaller than that of the conventional equal power allocation, further verifying the outage probability advantage of proposed power allocation.

\section{Conclusion}
In this paper, we considered a UAV assisted wireless system, where a BS transmits to its GU destination with the help of a UAV node. An analytical
expression of the outage probability was derived for the UAV relay transmission over Rician fading channels. We formulated an outage probability minimization problem under a constraint on the total transmit power of BS and UAV relay and obtained an optimal power allocation solution to minimize the derived outage probability. The conventional equal power allocation was also considered as a benchmark. Numerical results demonstrated that the proposed optimal power allocation scheme significantly performs better than the conventional power allocation method in terms of the outage probability. In the future, a possible extension of our work is to examine the joint optimization of power allocation and UAV altitude deployment.

\appendices
\section{Proof of Theorem 1}
\begin{IEEEproof}
We substitute ${P_u} = {P_t} - {P_s}$ into the ${P_{\textrm{out}}}$, for the transmit power of BS at which the system outage probility ${P_{\textrm{out}}}$ is minimized, substituting (21) to $\frac{{d{P_{\textrm{out}}}}}{{d{P_s}}} = 0$ we obtain
\begin{align} \label{f_11}
&\frac{d}{{d{P_s}}}\left[ {1 - {Q_1}\left( {\sqrt {2{K_{su}}} ,\sqrt {2({{2^{2R}} - 1})({K_{su}} + 1){N_0}/{P_s}{G_{su}}} } \right)} \right.\\ \notag
&\quad{\rm{      }}\left. { \times {Q_1}\left( {\sqrt {2{K_{ud}}} ,\sqrt {2({{2^{2R}} - 1})({K_{ud}} + 1){N_0}/{P_u}{G_{ud}}} } \right)} \right] = 0.
\end{align}

We define the auxiliary variables $\alpha$ and $\beta$  as
\begin{align} \label{f_13}
&{\alpha _1} \buildrel \Delta \over = \sqrt {2{K_{su}}}, \qquad   {\alpha _2} \buildrel \Delta \over = \sqrt {2{K_{ud}}}, \\
&{\beta _1} \buildrel \Delta \over = \sqrt {2{({{2^{2R}} - 1})}({K_{su}} + 1){N_0}/{P_s}{G_{su}}} {\rm{ ,}}  \\
&{\beta _2} \buildrel \Delta \over = \sqrt {2{({{2^{2R}} - 1})}({K_{ud}} + 1){N_0}/{P_u}{G_{ud}}} {\rm{ .}}
\end{align}

Then equation (24) can be simplified as
\begin{align} \label{f_14}
&\frac{{d{P_{\textrm{out}}}}}{{d{P_s}}} =  - \frac{{\partial {Q_1}({\alpha _1},{\beta _1})}}{{\partial {P_s}}}{Q_1}({\alpha _2},{\beta _2}) - \frac{{\partial {Q_1}({\alpha _2},{\beta _2})}}{{\partial {P_s}}}\\ \notag
&\qquad\qquad\qquad\qquad\qquad\qquad\qquad {\rm{                             }} \times {Q_1}({\alpha _1},{\beta _1}) = 0.
\end{align}

According to the derivation rules,
\begin{align} \label{f_15}
&\frac{{\partial {Q_1}({\alpha _1},{\beta _1})}}{{\partial {P_s}}} = \frac{{\partial {Q_1}({\alpha _1},{\beta _1})}}{{\partial {\alpha _1}}}\frac{{\partial {\alpha _1}}}{{\partial {P_s}}}{\rm{ + }}\frac{{\partial {Q_1}({\alpha _1},{\beta _1})}}{{\partial {\beta _1}}}\frac{{\partial {\beta _1}}}{{\partial {P_s}}} \\ \notag
&\frac{{\partial Q({\alpha _2},{\beta _2})}}{{\partial {P_s}}} = \frac{{\partial Q({\alpha _2},{\beta _2})}}{{\partial {\alpha _2}}}\frac{{\partial {\alpha _2}}}{{\partial \left( {{P_t} - {P_s}} \right)}} \\ \notag
&\qquad\qquad\qquad\qquad\qquad\quad {\rm{ + }}\frac{{\partial Q({\alpha _2},{\beta _2})}}{{\partial {\beta _2}}}\frac{{\partial {\beta _2}}}{{\partial \left( {{P_t} - {P_s}} \right)}}.
\end{align}
From [16] one can see that
\begin{align} \label{f_16}
&\frac{{\partial {Q_1}({\alpha _k},{\beta _k})}}{{\partial {\alpha _k}}}{\rm{ = }}{\beta _k}{e^{ - \frac{{{\alpha _k}^2 + {\beta _k}^2}}{2}}}{I_1}({\alpha _k}{\beta _k}),
\end{align}
\begin{align} \label{f_17}
&\frac{{\partial {Q_1}({\alpha _k},{\beta _k})}}{{\partial {\beta _k}}}{\rm{ =  - }}{\beta _k}{e^{ - \frac{{{\alpha _k}^2 + {\beta _k}^2}}{2}}}{I_0}({\alpha _k}{\beta _k}), \quad k = \left\{ {1,2} \right\}.
\end{align}

The partial derivatives of $\alpha$ and $\beta$ to ${P_s}$ are
\begin{align} \label{f_18}
&\frac{{\partial {\alpha _k}}}{{\partial {P_s}}} = 0,\ \frac{{\partial {\beta _1}}}{{\partial {P_s}}} =  - \frac{1}{2}\sqrt {\frac{{2({K_{su}} + 1){({{2^{2R}} - 1})}{N_0}}}{{{G_{su}}}}} {\left( {\frac{1}{{{P_s}}}} \right)^{\frac{3}{2}}},
\end{align}

\begin{align} \label{f_19}
&\frac{{\partial {\beta _2}}}{{\partial {P_s}}} = \frac{1}{2}\sqrt {\frac{{2({K_{ud}} + 1){({{2^{2R}} - 1})}{N_0}}}{{{G_{ud}}}}} {\left( {\frac{1}{{{P_t} - {P_s}}}} \right)^{\frac{3}{2}}}.
\end{align}

Then we obtain
\begin{align} \label{f_20}
&\frac{{{\beta _1}}}{{{\beta _2}}}\frac{{{e^{ - \frac{{{\alpha _1}^2 + {\beta _1}^2}}{2}}}}}{{{e^{ - \frac{{{\alpha _2}^2 + {\beta _2}^2}}{2}}}}}\frac{{{I_0}({\alpha _1}{\beta _1})}}{{{I_0}({\alpha _2}{\beta _2})}}\sqrt {\frac{{({K_{su}} + 1){G_{ud}}}}{{({K_{ud}} + 1){G_{su}}}}} {(\frac{{{P_t} - {P_s}}}{{{P_s}}})^{\frac{3}{2}}}\\ \notag
& \qquad\qquad\qquad\qquad\qquad\qquad\qquad\qquad {\rm{                             }} \times \frac{{{Q_1}({\alpha _2},{\beta _2})}}{{{Q_1}({\alpha _1},{\beta _1})}} = 1.
\end{align}

From [16] one can see that
\begin{align} \label{f_21}
&{Q_1}(\alpha, \beta ) = {e^{ - \frac{{{\alpha ^2} + {\beta ^2}}}{2}}}{\sum\limits_{n = 0}^\infty  {\left( {\frac{\alpha }{\beta }} \right)} ^n}{I_n}\left( {\alpha \beta } \right).
\end{align}

In order to solve this equation, we ignore the first few items of $n$ smaller than $x$ in the process of $n \to \infty$. For argument $x$ is small enough than $n$,we have[17]
\begin{align} \label{f_22}
&{I_n}\left( x \right) \sim {\left( {\frac{x}{2}} \right)^n}/\Gamma \left( {n + 1} \right).
\end{align}

Deriving the gamma function by partial integration has the following recursive properties
\begin{align} \label{f_24}
&\Gamma \left( {n + 1} \right) = n\Gamma \left( n \right).
\end{align}

It is easy to prove that the gamma function can be regarded as the extension of the factorial on the real set. For the positive integer, it has the following properties
\begin{align} \label{f_25}
&\Gamma \left( {n + 1} \right) = n!.
\end{align}

Thus we have
\begin{align} \label{f_26}
&\frac{{{G_{ud}}({K_{su}} + 1)}}{{{G_{su}}({K_{ud}} + 1)}}{(\frac{{{P_t} - {P_s}}}{{{P_s}}})^{\rm{2}}}\frac{{{I_0}\left( {{\alpha _1}{\beta _1}} \right)}}{{{I_0}\left( {{\alpha _2}{\beta _2}} \right)}}\frac{{\sum\limits_{n = 0}^\infty  {\frac{{{{\left( {{K_{ud}}} \right)}^n}}}{{n!}}} }}{{\sum\limits_{n = 0}^\infty  {\frac{{{{\left( {{K_{su}}} \right)}^n}}}{{n!}}} }} = 1.
\end{align}

For argument $x \gg n$, from [17] we obtain
\begin{align} \label{f_27}
&{I_n}(x) \cong \frac{{{e^x}}}{{\sqrt {2\pi x} }}\left( {1 - \frac{{n - 1}}{{8x}} + \frac{{\left( {n - 1} \right)\left( {n - 9} \right)}}{{2!{{\left( {8x} \right)}^2}}} -  \cdots } \right).
\end{align}

On this issue, we use the following approximation
\begin{align} \label{f_28}
&{I_0}\left( {\alpha \beta } \right) \cong \frac{{{e^{\alpha \beta }}}}{{\sqrt {2\pi \alpha \beta } }}.
\end{align}

According to the results in (23), finally we obtain
\begin{align} \label{f_28}
&{\left( {\frac{{{P_u}}}{{{P_s}}}} \right)^{\frac{7}{4}}}{\left( {\frac{{{\gamma _1}}}{{{\gamma _2}}}} \right)^{\frac{3}{4}}}\sqrt {\frac{{{K_{ud}}}}{{{K_{su}}}}} \exp [2\sqrt {\frac{{{\gamma _1}}}{{{P_s}}}}  - 2\sqrt {\frac{{{\gamma _2}}}{{{P_u}}}} \\
& \qquad\qquad\qquad\qquad\qquad\qquad + {K_{ud}} - {K_{su}}] = 1,\\ \notag
\end{align}
where \\
${\gamma _1} = \frac{{{K_{su}}({K_{su}} + 1){({{2^{2R}} - 1})}{N_0}}}{{{G_{su}}}}$, ${\rm{ }}{\gamma _2} = \frac{{{K_{ud}}({K_{ud}} + 1){({{2^{2R}} - 1})}{N_0}}}{{{G_{ud}}}}$.

The solution of the equation is satisfied $\frac{{d{P_{\textrm{out}}}}}{{d{P_s}}} = 0$. From the second-order sufficient condition of the minimum value, it is easy to prove $\frac{{{d^{\rm{2}}}{P_{\textrm{out}}}}}{{d{P_s}^{\rm{2}}}} > 0$. Therefore, the solution of the equation is the power allocation scheme that minimizes the outage probability.
\end{IEEEproof}

\end{document}